\documentclass[final,5p,times,twocolumn]{elsarticle}

\usepackage{graphicx}
\graphicspath{{figures/}}
\DeclareGraphicsExtensions{.pdf,.png}

\usepackage{amssymb}

\usepackage{amsmath}
\usepackage[colorlinks=true, citecolor=blue, linkcolor=blue]{hyperref}   
\usepackage{fancyhdr}
\journal{Journal of Quantitative Spectroscopy and Radiative Transfer}

\newcommand{\vect}[1]{\ensuremath{\mathbf{#1}}}
\newcommand{\xhat}[0]{\vect{\hat{x}}}
\newcommand{\yhat}[0]{\vect{\hat{y}}}

\begin{document}

\begin{frontmatter}



  \title{Using the Discrete Dipole Approximation and Holographic
    Microscopy to Measure Rotational Dynamics of Non-spherical
    Colloidal Particles}


\author[label1,*]{Anna Wang}
\author[label2,*]{Thomas G.~Dimiduk}
\author[label2]{Jerome Fung}
\author[label3]{Sepideh Razavi}
\author[label3] {Ilona Kretzschmar}
\author[label1] {Kundan Chaudhary}
\author[label1,label2]{Vinothan N.~Manoharan\corref{cor1}}
\ead{vnm@seas.harvard.edu}

\address[label1]{School of Engineering \& Applied Sciences, Harvard University, 29 Oxford Street, Cambridge, MA 02138, USA}
\address[label2]{Department of Physics, Harvard University, 17 Oxford Street, Cambridge, MA 02138, USA}
\address[label3]{Department of Chemical Engineering, City College of City University of New York, NY 10031, USA}
\fntext[*]{A. Wang and T.G. Dimiduk contributed equally to this work.}
\cortext[cor1]{Corresponding author}

\begin{abstract}
  We present a new, high-speed technique to track the
  three-dimensional translation and rotation of non-spherical
  colloidal particles.  We capture digital holograms of
  micrometer-scale silica rods and sub-micrometer-scale Janus
  particles freely diffusing in water, and then fit numerical
  scattering models based on the discrete dipole approximation to the
  measured holograms. This inverse-scattering approach allows us to
  extract the position and orientation of the particles as a
  function of time, along with static parameters including the size,
  shape, and refractive index. The best-fit sizes and refractive
  indices of both particles agree well with expected values. The
  technique is able to track the center of mass of the rod to a
  precision of 35 nm and its orientation to a precision of
  $1.5^\circ$, comparable to or better than the precision of other 3D
  diffusion measurements on non-spherical particles.  Furthermore, the
  measured translational and rotational diffusion coefficients for the
  silica rods agree with hydrodynamic predictions for a spherocylinder
  to within 0.3\%.  We also show that although the Janus particles
  have only weak optical asymmetry, the technique can track their 2D
  translation and azimuthal rotation over a depth of field of several
  micrometers, yielding independent measurements of the effective
  hydrodynamic radius that agree to within 0.2\%. The internal and
  external consistency of these measurements validate the technique.
  Because the discrete dipole approximation can model scattering from
  arbitrarily shaped particles, our technique could be used in a range
  of applications, including particle tracking, microrheology, and
  fundamental studies of colloidal self-assembly or microbial motion.
\end{abstract}


\end{frontmatter}


\section{Introduction}
\label{sec:intro}
\thispagestyle{fancy}

Measurements of the dynamics of colloidal particles are key to
understanding the mechanisms of colloidal
aggregation~\cite{weitz_dynamics_1984} and
self-assembly~\cite{glotzer_self-assembly_2004}.  Microscopic
measurements of the diffusion of individual particles can furthermore
be used to infer interactions between
particles~\cite{biancaniello_line_2006} as well as the local
rheological properties of the medium in which the particles are
suspended~\cite{chen_rheological_2003}. Most microscopic measurements
are constrained to two dimensions because of the limited depth of
field of wide-field microscopy~\cite{anthony_single-particle_2006}.
Confocal microscopy can be used to capture three-dimensional (3D)
dynamics of spherical and non-spherical
particles~\cite{hunter_tracking_2011, kraft_brownian_2013}, but the
time required to scan the beam through a 3D sample limits these
measurements to large particles or particles in a viscous fluid.

Holographic microscopy is an alternative technique that can in
principle capture 3D colloidal dynamics with acquisition times orders
of magnitude smaller than those of confocal microscopy.  In a
holographic microscope, light from a coherent source scatters from the
sample and interferes with a reference wave---which can simply be the
transmitted, undiffracted beam.  The interference pattern, or
hologram, contains phase information about the scattered wave and can
be used to determine the 3D position of the particle.  Because
holograms can be captured as fast as a camera allows and then
processed offline, the technique can be used to probe millisecond or
even microsecond dynamics. The 3D information is typically recovered
through optical or numerical
reconstruction~\cite{kreis_frequency_2002}.  Cheong and
co-workers~\cite{cheong_rotational_2010} used this method to track the
3D translation and rotation of a high-aspect-ratio (1:25) copper oxide
nanoparticle through numerical reconstruction and a skeletonization
procedure.  However, for wavelength-scale particles the precision of
reconstructions is limited owing to distortions in the reconstructed
volume~\cite{pu_intrinsic_2003, cheong_strategies_2010}.

We demonstrate a precise way to measure the dynamics of
wavelength-scale, non-spherical colloidal particles from holographic
measurements.  Our approach, which uses
an inverse-scattering analysis rather than reconstruction, follows the
work of Lee and coworkers~\cite{lee_characterizing_2007}, who showed
that fitting a Lorenz-Mie scattering solution to a measured hologram
yields nanometer-scale-precision measurements of the positions of
spherical particles. Researchers from our group, including Fung and
coworkers~\cite{fung_measuring_2011, fung_holographic_2013} and Perry
and coworkers~\cite{perry_real-space_2012} later showed that the
translational, rotational, vibrational, and nonequilibrium dynamics of
several interacting wavelength-scale spheres could be measured with
high precision by fitting Mackowski and Mishchenko's T-matrix
solution~\cite{mackowski_calculation_1996} to holograms of multiple
colloidal particles.  All of these measurements were limited to
spheres, or collections of spheres, because the fitting technique
requires a solution for the scattered field, and exact solutions are
known for very few particle morphologies.

One can, however, perform approximate numerical scattering
calculations for a wide variety of particles using the Discrete Dipole
Approximation (DDA) of Purcell and
Pennypacker~\cite{purcell_scattering_1973}.  We show that numerical
scattering calculations using the DDA can be fit to holograms of
non-spherical particles, allowing us to track 3D translational and
rotational dynamics at high precision, even for particles with subtle
asymmetries. We use these measurements to determine translational and
rotational diffusion coefficients and show that the measured values
agree well with theoretical calculations, validating the technique.
Although DDA calculations are orders of magnitude slower than
calculations of exact scattering solutions, parallelization of the
scattering calculations and the fitting algorithm can significantly
reduce the analysis time.

\begin{figure}[htbp]
\centering
\includegraphics{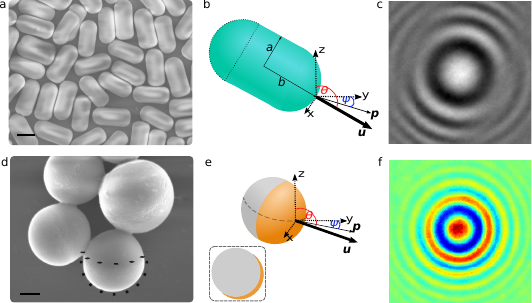}
\caption{\label{fig:particles} The non-spherical particles used in our
  experiments. Top row, silica rods; bottom row, Janus particles
  consisting of a polystyrene sphere coated with a 50-nm-thick layer
  of TiO$_2$. The orientation of both types of particles is defined by
  the orientational unit vector $\mathbf{u}$ and its projection onto
  the $x-y$ plane, $\mathbf{p}$. The z-axis coincides with the
  direction of the incident plane wave illumination. (a) Scanning
  electron micrograph of the rods. Scale bar is 1 $\mu$m. (b) The rod
  is modeled as a spherocylinder with semi-minor-axis length $a$
  and semi-major-axis length $b$. (c) Measured hologram of a
  single rod. (d) Scanning electron micrograph of the Janus
  particles. The dotted line outlines the TiO$_2$ cap. Scale bar is
  300 nm. (e) The Janus particle is modeled as a plain polystyrene
  sphere with a hemispherical cap of TiO$_2$ (orange). A cross section
  perpendicular to the equator is also shown. (f) Measured hologram of
  a Janus particle, shown in false color to highlight asymmetry.}
\end{figure}

\section{Experimental Methods}
\label{sec:methods}

We use two types of anisotropic particles in this study: silica rods
and polystyrene/TiO$_2$ Janus particles suspended in water
(Figure~\ref{fig:particles}). The silica rods are synthesized using a
modified one-pot method recently reported by Kuijk and
coworkers~\cite{chaudhary_janus_2012, kuijk_synthesis_2011}. In the
initial growth step, we make silica rods with a length (L) of $1.45
\pm 0.06$ $\mu$m and diameter (D) of $0.29 \pm 0.02$ $\mu$m, which are
then used as seeds in subsequent growth
steps~\cite{stober_controlled_1968} to produce the final silica
rods. The size of the rods, as determined by scanning electron
microscopy (SEM), is $1.0 \pm 0.2$ $\mu$m (minor axis) by $2.0 \pm 0.2$
$\mu$m (major axis) (Figure~\ref{fig:particles}a). We fabricate the
Janus particles by depositing 50 nm of TiO$_2$ onto sulfate-terminated
polystyrene (PS) particles (Invitrogen), as described in
reference~\cite{song_assembled_2009}. The polystyrene particle
diameter, as determined by SEM, is $900 \pm 100$ nm (neglecting
anomalously large particles). The TiO$_2$ cap covers approximately one
hemisphere of the particle, as shown in Figure~\ref{fig:particles}d.

For holographic imaging, we suspend the particles in deionized water
at approximately $10^{-5}$ volume fraction and place them in sample
cells consisting of two No.\,1 glass coverslips (VWR) separated and
sealed by vacuum grease (Dow Corning).  To validate the DDA method, we
also image a 0.95-$\mu$m-radius polystyrene sphere (Invitrogen)
diffusing in a 54\% v/v glycerol solution.

\begin{figure}[htbp]
\centering
\includegraphics{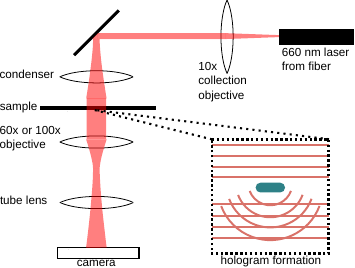}
\caption{\label{fig:Uberscope} Schematic of our digital in-line
  holographic microscope. A series of lenses, including the condenser,
  shapes light from a 660 nm laser diode (Opnext HL6545MG) into a
  plane wave to illuminate the sample. An objective collects the
  transmitted light and light scattered from the sample, and a
  high-speed camera (Photon Focus) captures the hologram formed by the
  interference between scattered and transmitted
  waves.}
\end{figure}

We capture holograms with a digital holographic microscope built on a
Nikon (TE2000-E) inverted microscope, as shown in
Figure~\ref{fig:Uberscope} and described in detail
in~\cite{fung_measuring_2011, kaz_physical_2012}.  When imaging the
Janus particles we use a 100$\times$ oil-immersion objective (Nikon)
with a high numerical aperture (1.40) that allows us to maximize
detail in the holograms.  When imaging the silica rods, which sediment
rapidly owing to their density, we use a 60$\times$ water-immersion
objective (Nikon), which has a lower numerical aperture (1.20) but a
larger working distance that allows us to capture longer trajectories.
To obtain true bulk-diffusion measurements, we retain data only from
particles that remain at least 20 $\mu$m away from the sample cell
boundaries throughout the trajectory.

In a typical experiment we use 50 mW of laser power and a 0.015 ms
exposure time per camera frame for the 60$\times$ lens or 0.05 ms for
the 100$\times$ lens. These exposure times are short enough to
minimize blurring due to Brownian motion. We capture holograms with a
Photon Focus MVD-1024E-160 camera at 100 frames per second, store them
in RAM using a frame grabber (EPIX PIXCI E4), and then save to disk
for further processing. For each trajectory we also record a
background image from the same region of the sample cell before or
after the particle is present to account for scattering and
illumination defects in the optical train.

\section{Fitting holograms using the DDA}

In contrast to holograms of spheres, holograms of non-spherical
particles do not have azimuthal symmetry about the central maximum
(see Figure~\ref{fig:particles}c, f).  The azimuthal asymmetry encodes
information about the particle shape and orientation, while the radial
spacing of the interference fringes encodes the particle position.
To quantitatively extract this information, we fit a scattering model
to the holograms.

In our fitting procedure, we compute holograms from a model, compare
the computed holograms to experimentally recorded ones, and iterate
until the model matches the data. To compute the hologram, we model
the physical process of hologram formation: the interference of
scattered and reference fields.  In an inline hologram, the reference
field is approximately equal to the incident field $\mathbf{E}_i$, as
long as the density of scatterers is low. In our experiments, we work
at low concentrations of particles, so that we can assume
$\mathbf{E}_i$ is a constant plane wave. The observed intensity is
therefore
\begin{equation}
  \label{eq:holo}
  I_\mathrm{holo} = |\mathbf{E}_i + \mathbf{E}_s|^2,
\end{equation}
where $\mathbf{E}_s$ is the scattered field, which we must compute to
simulate the hologram. Previous work~\cite{lee_characterizing_2007,
  fung_measuring_2011, perry_real-space_2012, kaz_physical_2012} used
numerical implementations of the Lorenz-Mie solution or a multi-sphere
superposition solution~\cite{mackowski_calculation_1996} to compute
$\mathbf{E}_s$. However, these exact solutions are limited to spheres
or collections of spheres. For the non-spherical particles used here,
we must compute $\mathbf{E}_s$ approximately.  We do this using the
discrete dipole approximation, as implemented by the open-source
scattering code ADDA~\cite{yurkin_discrete-dipole-approximation_2011}.

\begin{figure}[htbp]
\centering
\includegraphics{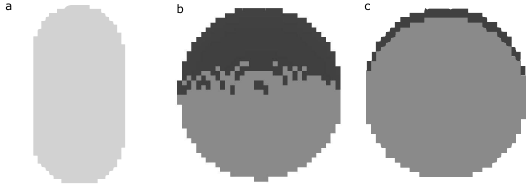}
\caption{\label{fig:voxel} Voxelations of the particles. White
  represents vacuum, light gray is silica, medium gray is polystyrene
  and dark gray is TiO$_2$. Each dipole is represented by a
  square. (a) Side view of a rod with a dipole size of 44 nm, a
  semi-minor-axis length of 500 nm and a semi-major-axis length of 1
  $\mu$m.  (b) Side view of a Janus particle with a dipole size of 24
  nm, inner particle radius of 439 nm and a maximum cap thickness of
  50 nm.  (c) Cross section of the same Janus particle.}
\end{figure}

In the DDA, a scatterer is represented as an array of point dipoles.
Therefore the first step of our data analysis is to discretize the
particle, dividing it into an array of voxels (volumetric pixels). To
ensure validity of the DDA, we discretize the scatterer using at least
10 dipoles per wavelength in the scattering medium, as recommended by
the ADDA
documentation~\cite{yurkin_discrete-dipole-approximation_2011}. The
voxelations of both particles are shown in Figure \ref{fig:voxel}.
For the Janus particle, the dipole size is comparable to the thickness
of the TiO$_2$ layer; but because the shape and size of the cap varies
from particle to particle, a smaller dipole size would not necessarily
increase accuracy.

To compute a hologram, we calculate the scattering angles for each
pixel in the detector, then invoke ADDA to compute the scattering
matrix of the voxelated particle at each scattering angle. We obtain
the electric field $\mathbf{E}_s(i,j)$ at each pixel $(i,j)$ on the
detector from the scattering matrices, angles, and distances.
Finally, we numerically interfere the computed $\mathbf{E}_s(i,j)$
with $\mathbf{E}_i(i,j)$, a plane wave, to obtain the hologram.  All
of these steps are implemented in our open-source hologram processing
code, HoloPy
(\href{http://manoharan.seas.harvard.edu/holopy/}{http://manoharan.seas.harvard.edu/holopy/}).

Once we have created a DDA scattering model for a particle, we use
nonlinear minimization to fit the model to measured,
back\-ground-divided, normalized holograms, thereby obtaining
measurements of the positions, refractive index, and geometrical
parameters of individual particles.  Formally, we use the
Levenberg-Marquardt algorithm to minimize the objective function
\begin{equation}
\label{eq:minimize}
  f(\{p\}) = \sum_{i,j}\left|I_\mathrm{measured}(i,j) - I_\mathrm{computed}\left(i,j; \{p\}\right) \right|^2 = \chi^2
\end{equation}
where
\begin{equation*}
I_\mathrm{measured} = \frac{I_\mathrm{data}}{I_\mathrm{background}}\frac{\overline{I}_\mathrm{background}}{\overline{I}_\mathrm{data}},
\end{equation*}
and $\{p\}$ is the set or a subset of the parameters used by the DDA
model to calculate the scattering from the particle.  The minimization
typically requires several iterations to converge.  We then use the
fitted position and orientations obtained from a time-series of
holograms to compute the diffusion coefficients.

Because the two types of particles have different geometries, we must
use different sets of parameters to model them. We model the rod,
which has rounded ends as shown in Figure \ref{fig:particles}a, as a
dielectric spherocylinder with refractive index $n$, hemisphere radius
$a$ (equal to the length of the semi-minor-axis), and semi-major-axis
length $b$.  The full set of parameters is
\begin{equation}
  \{p_\mathrm{rod}\} = \{\mathbf{r}, n, a, b, \theta, \psi, \alpha\},
\end{equation}
where $\mathbf{r}$ is the center of mass position, $\theta$ and $\psi$
are Euler angles describing particle orientation (see
Figure~\ref{fig:particles}b), and $\alpha$ is a normalization
constant~\cite{lee_characterizing_2007}. We model the Janus particle
(Figure \ref{fig:particles}d) as a dielectric sphere with refractive
index $n_\mathrm{PS}$ and radius $a_\mathrm{Janus}$ capped by a
hemispherical dielectric shell with index $n_\mathrm{TiO_2}$ and
thickness $t$.  Here the set of parameters is
\begin{equation}
  \{p_\mathrm{Janus}\} = \{\mathbf{r}, n_\mathrm{PS}, n_\mathrm{TiO_2}, a_\mathrm{Janus}, t, \theta, \psi, \alpha\},
\end{equation}
where $\mathbf{r}$ is the center of the sphere, $\theta$ and $\psi$
are the Euler angles shown in Figure \ref{fig:particles}e, and
$\alpha$ is again a normalization constant.

\begin{figure*}[htbp]
\centering
\includegraphics{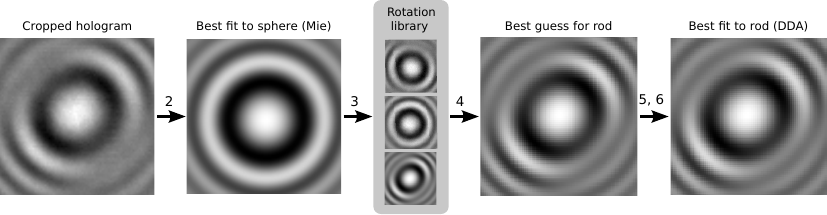}
\caption{\label{fig:initialguess} Procedure for fitting DDA models to
  experimental holograms of non-spherical particles to extract
  position and orientation. The numbers refer to the steps in the
  text.}
\end{figure*}

To assess the validity of the fits, we examine the coefficient of
determination, $R^2$, which measures how much of the variation from
the mean value of one hologram is captured in another:
\begin{equation}
  R^2 = 1 - \frac{\sum_{i,j}\left|I_{1}(i,j) - I_{2}(i,j)\right|^2}{\sum_{i,j} \left|I_{1}(i,j) - \bar{I}_{1}\right|^2},
\end{equation}
where $\bar{I}_{1}$ = 1 is the mean value of the normalized
hologram. $R^2$ is a useful quantity because unlike $\chi^2$, which is
sensitive to the amplitude of the hologram fringes, the $R^2$ values
for good fits do not vary markedly across experimental systems with
different scattering cross sections~\cite{fung_imaging_2012}. We use
$R^2$ to assess how well DDA reproduces the results of exact
Lorenz-Mie calculations for holograms of spheres and to assess how
well the holograms from our model match data from experiments.

Because DDA calculations take at least an order of magnitude longer
than Lorenz-Mie calculations for scatterers of the same size, we aim
to fit all the frames in parallel.  This goal is complicated by
convergence requirements.  In practice, we have found that the
minimization does not converge unless the initial guess for $\{p\}$ is
close to the global minimum of the objective function.  Therefore our
usual procedure~\cite{fung_measuring_2011, fung_holographic_2013} is
to fit the frames sequentially: we use reconstruction to manually
obtain an initial guess for the first frame of the trajectory and then
use the best-fit result for each frame as the initial guess for the
next. Here such a process would take too long. A DDA calculation takes
on the order of 30 seconds on a single CPU; since about 200 such
calculations are required for each minimization to converge, it would
take 20 weeks to sequentially fit a 2000-frame trajectory.  We
therefore use the procedure described below and illustrated in
Figure~\ref{fig:initialguess} to obtain the initial guesses for each
frame and fit the frames in parallel:

\begin{enumerate}
\label{list:init}
\item \emph{Select the most asymmetric region:} The particle
  orientation is primarily encoded in the rotational asymmetry of a
  hologram. We maximize sensitivity to the particle orientation by
  cropping the hologram to leave only the most asymmetric region, the
  first fringe or two around the central maximum. We use a Hough
  transform-based algorithm~\cite{cheong_flow_2010} to locate the
  center of the hologram. For our experiments the first fringe is
  located within a $140 \times 140$ pixel area.

\item \label{item:miefit} \emph{Find approximate position with a
    Lorenz-Mie fit:} We obtain an estimate of the particle's position
  by approximating it as a sphere with parameters $\{p\} =
  \{\mathbf{r_\mathrm{Mie}}, n, a_\mathrm{Mie}, \alpha\}$ and fitting
  the Lorenz-Mie solution to the hologram. This fit is very fast
  (order of 10 seconds per frame) and although $\chi^2$ is large,
  $\mathbf{r}_\mathrm{Mie}$ provides a good initial guess for the
  position of the particle's center.
\item \emph{Make rotation library:} We use the DDA to
  calculate holograms of the non-spherical particle over a range of
  Euler angles $\theta$, $\psi$, assuming a center position
  $\mathbf{r}_\mathrm{Mie}$.
\item \label{item:rot_guess} \emph{Use library to guess angles:} We
  compare this library to the experimental hologram and determine the
  Euler angles $\theta$, $\psi$ that yield the highest $R^2$ value.
\item \label{item:dda_initial}\emph{Initial DDA fit:} We fit the DDA
  model to the measured hologram using the guess for
  $\mathbf{r}_\mathrm{Mie}$ from step \ref{item:miefit}, $\theta$,
  $\psi$ from step \ref{item:rot_guess}, bulk values for the index or
  indices, and an estimate of the geometrical parameters $a$ and $b$
  or $a_\mathrm{Janus}$ and $t$ from SEM images.  We allow all parameters
  to vary during the fit. In this step, we fit the first 500 frames of
  a trajectory, discard the fits that fail to converge, then
  calculate the average best-fit refractive indices and radii.
\item \label{item:dda_final} \emph{Final DDA fit:} Because the
  particle's physical properties should not change across a
  trajectory, we fix the refractive indices and radii to the average
  values from step \ref{item:dda_initial} and fit the cropped
  holograms for an entire trajectory to obtain the position and
  orientation as a function of time.
\end{enumerate}

This procedure removes the serial dependency between the frames in a
time series, allowing us to send each frame in the initial
(step~\ref{item:dda_initial}) and final (step~\ref{item:dda_final})
DDA fits to an individual CPU on a computing cluster.  However,
removing the serial dependency potentially introduces artifacts in the
trajectories of the particles, since the results from each frame are
not used to constrain the results of the next.  We must therefore
detect and correct spurious fit results.

There are two kinds of errors. The first arises because the rod
particles are symmetric, having two degenerate orientational unit
vectors pointing in opposite directions. In a trajectory constructed
from a series of parallel fits, the fitted orientational unit vector
may therefore flip almost 180$^{\circ}$ between frames.  Using the
probability density function for rotational displacements described in
reference~\cite{fung_holographic_2013} and an estimated rotational
diffusion coefficient $D_\mathrm{est} = 0.3$ s$^{-1}$, we estimate
that the probability of a 90$^{\circ}$ or greater change in angle
between frames is zero, to within machine precision.  Therefore we can
correct artificial flips by examining pairs of frames in the
trajectory and flipping the vector in the second frame by
180$^{\circ}$ if the angle changes by more than 90$^{\circ}$ between
frames.

The second type of spurious fit result arises because the
Levenberg-Marquardt algorithm is sensitive to noise.  Depending on the
initial guess, a fit can converge to a secondary minimum in which the
orientational unit vector $\vect{u}$ points in a direction mirrored
about the x-y plane from that of the previous frame.  A small number
of these bad fits systematically increases the apparent diffusion
coefficient.

We therefore reject contributions to the mean-square displacement
where the probability of the measured angular displacement is less
than 0.0001. Again, we calculate the probability of observing a given
displacement using the probability density function for rotational
displacements described in reference~\cite{fung_holographic_2013} and
an estimated rotational diffusion coefficient $D_\mathrm{est} = 0.3$
s$^{-1}$. Using $D_\mathrm{est} = 0.2$ s$^{-1}$ or $D_\mathrm{est} =
0.4$ s$^{-1}$ changes the coefficients extracted from the data by less
than 7\%, an effect which we include in the error on the measured
diffusion coefficients. This procedure eliminates poor fit results
while minimizing the bias of results toward our estimated diffusion
coefficient.

\section{Results and Discussion}
\label{sec:results}

We demonstrate the validity and usefulness of the DDA as a tool for
holography and particle tracking through a series of experimental and
computational tests. First we compare the results of DDA and
Lorenz-Mie calculations for scattering from spherical particles, which
the Lorenz-Mie solution models exactly.  Next we fit the DDA model to
holograms of a rod and a Janus particle, and we obtain translational
and rotational diffusion coefficients from the mean-square
displacements measured across time-series of holograms. We compare the
measured diffusion coefficients to predicted values to demonstrate
that we accurately and precisely capture the particle motion.

\subsection{Validation of DDA calculations and fits}
\label{sec:nearfield}

We first examine how the approximations involved in our DDA
implementation affect the quality of the fits and calculated
holograms.  To do this, we compare holograms of spheres calculated
with our DDA model to those calculated from the Lorenz-Mie solution.
We are interested in two effects: first, how aliasing introduced by
the voxelation affects the precision of fitting holograms and, second,
how the absence of a near-field calculation in ADDA affects the
accuracy of the holograms.  By ``near field'' we mean the part of the
scattered field that does not follow an asymptotic $1/r$ decay.  ADDA
calculates only the asymptotic portion of the scattered field.  In
previous work~\cite{fung_measuring_2011, fung_imaging_2012} we found
that it was necessary to include the full radial dependence of
$\mathbf{E}_s$ to accurately fit holograms of particles close to the
focal plane.

We test the validity of the scattering model, including our voxelation
scheme, by fitting the DDA model and the Lorenz-Mie solution to 893
measured holograms of a 0.95-$\mu$m-radius polystyrene sphere
(Invitrogen) freely diffusing in a 54\% v/v glycerol solution. We fix
both the refractive index and radius of the particle and allow only
the center position of the sphere to vary.  Throughout the trajectory,
the particle remains 15 $\mu$m to 18 $\mu$m from the focal plane.

We find that the best-fit $x$- and $y$-coordinates obtained by fitting the
DDA model differ negligibly ($0.3 \pm 0.1$ nm) from those obtained
from Lorenz-Mie, while the best-fit $z$-coordinates differ by $194 \pm
1$ nm, a significant offset.  This difference in apparent axial
position might arise from voxelation errors or the asymptotic
approximation for the radial dependence of the scattered field. The
Lorenz-Mie model includes non-asymptotic corrections for the
scattered field, while the DDA model does not.

\begin{figure}[htbp]
\centering
\includegraphics{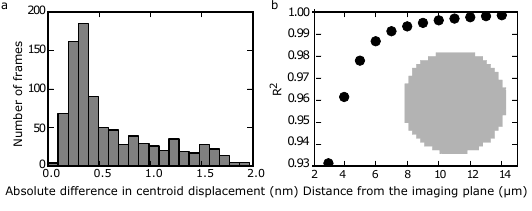}
\caption{\label{fig:ddamie} Comparison of the DDA model and
  Lorenz-Mie solution for fitting holograms of a single sphere. (a)
  Histogram of differences between displacements of a 950-nm-radius
  diffusing sphere measured by fitting the DDA and Lorenz-Mie models
  to measured holograms. (b) $R^2$ for pairs of holograms calculated
  from the exact Lorenz-Mie solution and the DDA model of a
  450-nm-radius sphere at various distances from the imaging
  plane. Inset: voxelation of the 450-nm-radius sphere used in the DDA
  model. The voxel size is 34 nm. The particle refractive index is
  1.585.}
\end{figure}

The offset in the $z$-coordinate does not affect our dynamical
measurements, which depend only on the displacement of the particle,
not its absolute position.  We find that the total displacements
$\|\mathbf{r}(t) - \mathbf{r}(0)\|$ for the DDA and Lorenz-Mie models,
as calculated from the best-fit coordinates for the entire trajectory,
agree to within 0.5 nm (Figure~\ref{fig:ddamie}a) for most of the
frames analyzed.  The maximum deviation in displacement is 2 nm, which
is within the reported accuracy of holographic measurements based on
inverse-scattering analysis~\cite{lee_characterizing_2007}.  Thus the
systematic error in the absolute particle position appears to cancel
when calculating the displacement, and the resulting displacement
measurements are nearly as precise as those obtained by fitting the
data to the exact Lorenz-Mie solution.

To further examine the impact of the asymptotic approximation for the
scattered field, we compare holograms computed from the DDA model and
Lorenz-Mie solutions for a 450-nm-radius sphere as a function of
distance from the focal plane. The agreement between the two models,
as measured by the $R^2$ value, increases with distance from the
imaging plane, as shown in Figure~\ref{fig:ddamie}b.  These data
suggest that near-field effects start to vary significantly with
distance when the particle is 5 $\mu$m or less from the focal
plane. To ensure the validity of our DDA fits, we take all
measurements with the particle at least 5 $\mu$m from the
imaging plane.

As another check on the validity of the technique, we examine the
average of fitted values for the optical properties and sizes of our
non-spherical particles.  We find $n = 1.495 \pm 0.012$ for the rod,
$n_\mathrm{PS} = 1.581 \pm 0.046$ for the polystyrene particle and
$n_\mathrm{TiO_2} = 2.74 \pm 0.23$ for the shell of the Janus
particle, all of which are close to bulk values for the materials. The
best-fit dimensions of the rod are $2a = 1.002 \pm 0.037$ $\mu$m and
$2b = 2.158 \pm 0.143$ $\mu$m, in agreement with the measurements from
SEM. The fitted dimensions of a Janus particle are $a_\mathrm{Janus} = 443
\pm 20$ nm, in excellent agreement with the SEM images, and $t = 47
\pm 8$ nm, in good agreement with the expected value based on the
TiO$_2$ deposition process.

\begin{figure}[htbp]
\centering
\includegraphics{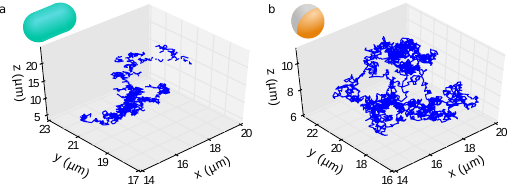}
\caption{\label{fig:diffusingparticles} Trajectories of particles as
  determined from DDA fits. (a) Rod trajectory captured at 100 frames
  per second for 20 seconds. (b) Janus particle trajectory captured at
  100 frames per second for 30 seconds.}
\end{figure}

Qualitatively, the trajectories of the particles we obtain by fitting
the DDA model to time-series of holograms are consistent with
stochastic Brownian motion
(Figure~\ref{fig:diffusingparticles}). Furthermore, the rod appears to
sediment at 0.9 $\mu$m/s whereas the Janus particle diffuses about a
steady height. This behavior agrees qualitatively with our
expectations for the two systems: the rod should sediment more rapidly
because the density difference between the rod and solvent is about
three times larger than that of the Janus particle, and the rod has
about three times the volume of the Janus particle.

Finally, to quantitatively investigate the validity of the DDA fits,
we determine the rotational and translational diffusion coefficients
from the data and compare them to theoretically predicted values, as
described in the following sections.

\subsection{Rods}
\label{sec:cylinderresult}

When viewed in the laboratory frame, the motion of a rod is
complicated because the viscous drag along any laboratory axis depends
on the orientation of the particle. But in the frame of the particle,
translational diffusion can be broken down into two components, one
parallel to the long axis of the particle ($\parallel$) and the other
along either of the two principal axes perpendicular to the long axis
($\perp$). Due to the optical symmetry of the particle, we cannot
observe rotations about the major axis.  We can therefore measure only
one rotational diffusion coefficient, corresponding to rotation about
either of the two degenerate principal axes that are coincident with
the minor axis.

To measure these diffusion coefficients, we use the best-fit positions
obtained from the holograms to calculate the translational and angular
mean-square displacements in the particle frame as a function of lag
time. We average these displacements over time and calculate the
uncertainties associated with these averages using a block
decorrelation method~\cite{flyvbjerg_error_1989}. Finally, we fit the
mean-square displacements using the calculated uncertainties as
weights. The uncertainties are represented as error bars in the
figures. In the following sections we report the formulas used
to calculate and fit the displacements, as well as our measured and
predicted values for the diffusion coefficients.

\subsubsection{Theoretical predictions}

For clarity, we denote all predicted values of diffusion coefficients
with a prime symbol ($'$). The diffusion coefficients of a
spherocylinder can be expressed as power series in the aspect ratio
$\omega$ = $b/a$, as discussed in the appendix of
reference~\cite{martchenko_hydrodynamic_2011}, which is adapted from
references~\cite{yoshizaki_dynamics_1980}
and~\cite{norisuye_wormlike_1979}:
\begin{multline}
  D'_t \approx \frac{k_B T}{6 \pi \eta b}
  \bigl(\ln\omega+0.3863+0.6863/\omega-0.0625/\omega^2-
  \\0.01042/\omega^3-0.000651/\omega^4+0.0005859/\omega^5\bigr)
\end{multline}
where $D_t = (2D_\perp + D_{\parallel})/3$.  We compare our results to
the predicted value for $D_t$, which is a linear combination of the
measured diffusion coefficients. From the same reference, we obtain
the rotational diffusion coefficient:
\begin{multline}
D'_r \approx \frac{3 k_B T}{8 \pi \eta b^3} \biggl(\ln\omega+2 \ln
2-\frac{11}{6}+ \\\frac{\ln2}{\ln(1+\omega)}\biggl[\frac{1}{3} -
2\ln2+\frac{11}{6}-|\vect{a}|\biggr]+
\vect{a}\cdot\vect{\Omega}\biggr)
\end{multline}
where
\begin{multline*}
\vect{a}=[13.04468, -62.6084, 174.0921, -218.8365, \\
140.26992, -33.27076],
\end{multline*}
and
\begin{equation*}
\vect{\Omega} = [\omega^{-1/4},\omega^{-2/4}, \omega^{-3/4},\omega^{-4/4},\omega^{-5/4},\omega^{-6/4}].
\end{equation*}

Because the rods are slightly asymmetric---flatter on one end of the
rod than the other (see Figure~\ref{fig:particles}a)---we also
compare our results for the translational and rotational diffusion
coefficients to theoretical predictions for a cylinder
with flat ends~\cite{tirado_comparison_1984}:
\begin{equation}
  D'_{\perp} \approx \frac{k_B T}{8 \pi \eta b}\left(\ln\omega+0.839+0.185/\omega+0.233/\omega^2\right)
\end{equation}
\begin{equation}
\label{t2}
  D'_{\parallel} \approx \frac{k_B T}{4 \pi \eta b}\left(\ln\omega-0.207+0.980/\omega-0.133/\omega^2\right)
\end{equation}
\begin{equation}
  D'_r \approx \frac{3 k_B T}{8 \pi \eta b^3}\left(\ln\omega-0.662+0.917/\omega-0.050/\omega^2\right).
\end{equation}

We use the laboratory temperature and uncertainty, $21\pm 2^\circ$C,
the corresponding viscosity for water, and the dimensions of the
particle from the hologram fits to calculate the predicted diffusion
coefficients and their uncertainties. As the values of rod's fitted
dimensions are not normally distributed, we do not include them when
propagating the errors.

\subsubsection{Translational diffusion}

To extract translational diffusion coefficients from the data, we
first calculate the mean-square displacements parallel and
perpendicular to the major axis.  For short lag times $\tau$ we
calculate
\begin{equation}
  \Delta {r^2}_{\parallel}(\tau) = \left\langle \left(\left(\textbf{r}(t + \tau)
    - \textbf{r}(t)\right) \cdot \textbf{u}(t)\right)^2 \right\rangle =
  2D_{\parallel}\tau + 2\epsilon_{\parallel}^2
\end{equation}
\begin{equation}
  \Delta {r^2}_{\perp}(\tau) = \left\langle |(\textbf{r}(t + \tau)
    - \textbf{r}(t)) \times \textbf{u}(t)|^2 \right\rangle =
  4D_{\perp}\tau + 4\epsilon_{\perp}^2
\end{equation}
where the angle brackets denote a time average over all contributing
pairs from a single trajectory. We then determine the diffusion
coefficients $D_\parallel$ and $D_\perp$ and measurement errors
$\epsilon_{\parallel}$ and $\epsilon_{\perp}$ by fitting a linear
model to the measured mean-square displacements.

\begin{figure}[htbp]
\centering
\includegraphics{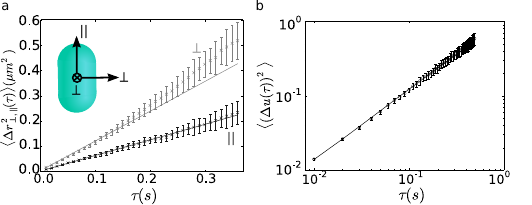}
\caption{\label{fig:cylinderresult} Measured mean-square displacements
  for the silica rods. (a) Mean-square displacements parallel and
  perpendicular to the major axis. The solid lines are fits to
  $2D_{\parallel} \tau + 2\epsilon_{\parallel}^2$ and $4D_{\perp} \tau
  + 4\epsilon_{\perp}^2$. (b) Mean-square displacement of the
  orientational unit vector $\mathbf{u}$. The solid line is a fit to
  the function $2-2\exp(-2D_r \tau) + 2\epsilon_r^2$.}
\end{figure}

Our measured results agree with those predicted by the
spherocylinder model.  We obtain $D_{\parallel} = 0.307 \pm 0.028$
$\mu$m$^2$/s, $D_{\perp} = 0.293 \pm 0.025$ $\mu$m$^2$/s and $D_t =
0.298 \pm 0.026$ $\mu$m$^2$/s (see Figure~\ref{fig:cylinderresult} and
Table~\ref{tab:rod_results}). These values are larger than the
predictions for a cylinder with flat ends ($D'_{\parallel} = 0.301 \pm
0.013$ $\mu$m$^2$/s and $D'_{\perp} = 0.265 \pm 0.012$ $\mu$m$^2$/s)
but agree quantitatively with the spherocylinder model ($D'_t = 0.297
\pm 0.016$ $\mu$m$^2$/s), suggesting that the rods are better modeled
as spherocylinders.

The fit, which is constrained by the small uncertainties at shorter
lag times, falls outside the error range for the perpendicular
direction at larger lag times (Figure~\ref{fig:cylinderresult}).
However, the fit for the parallel component is well within the error
bars throughout the entire range of lag times.  The apparent augmented
motion in the perpendicular direction may be due to radiation pressure
from the incident beam or coupling between sedimentation and
diffusion~\cite{brenner_taylor_1979}.

\begin{table*}[htbp]
  \centering
  \begin{tabular}{ l l l l}
 Quantity & Experiment & Expected & Source \\ \hline
 $a$ (nm) & $501 \pm 19$ & $500 \pm 50$ & SEM\\
 $b$ (nm) & $1079 \pm 72$ & $1000 \pm 100$ & SEM\\
 $n$ & $1.495 \pm 0.012$ & $\le1.54$ & bulk value \\
 $D_t$ ($\times 10^{-13}$ m$^2$s$^{-1}$) & $0.298 \pm 0.026$ & $0.297 \pm 0.016$ & spherocylinder model \\
 $D_r$ (s$^{-1}$) & $0.311 \pm 0.034$ &
 $0.311 \pm 0.017$ & spherocylinder model\\
 $D_{\parallel}/D_\perp$ & $1.04 \pm 0.05$ &
 1.13 & cylinder model\\
\end{tabular}
\caption{\label{tab:rod_results} Measured and expected values of parameters for silica rods.  Theoretical predictions for the spherocylinder model are based on relations in reference~\cite{martchenko_hydrodynamic_2011}, and those for the cylinder model on relations in reference~\cite{tirado_comparison_1984}.}
\end{table*}

The measured ratio $D_{\parallel}/D_{\perp} = 1.05 \pm 0.04$ depends
only on the aspect ratio of the particle.  Our value is smaller than
that predicted by the cylinder model ($D'_{\parallel}/D'_{\perp} =
1.13$). The small discrepancy in the measured and predicted values of
$D_{\parallel}/D_{\perp}$ is not surprising, given how much the
measured $D_\parallel$ and $D_\perp$ values differ individually from
those predicted by the cylinder model.  We are not able to calculate
$D_{\parallel}/D_{\perp}$ for the spherocylinder model because we do
not have explicit expressions for the two individual components.  In
general, however, we expect $D_{\parallel}/D_{\perp}$ to be of order
unity for particles that have an aspect ratio of 1:2.  Therefore our
measured value is physically reasonable.

From the linear fit to the mean-square displacement, we find that the
precision to which we track the rod's center of mass is 35 nm or
better ($\epsilon_{\parallel} = 28.3$ nm and $\epsilon_{\perp} = 35.4$
nm), which is around the same size as one voxel (approximately 35
nm). Though about an order of magnitude lower than the precision to
which single spheres can be tracked, this precision compares favorably
with that of tracking more complex scatterers, such as spheres in
clusters~\cite{fung_measuring_2011, perry_real-space_2012}.

\subsubsection{Rotational diffusion}
To determine the rotational diffusion coefficient, we first calculate
the mean-square displacement of the orientational unit vector
$\mathbf{u}$:
\begin{equation}
  \Delta\vect{u}^2(\tau) = \left\langle \left(\mathbf{u}(t+\tau) - \mathbf{u}(t)\right)^2 \right\rangle = 2\left(1-e^{-2D_r\tau}\right)+ 2\epsilon_r^2.
\end{equation}
We then fit $\Delta\vect{u}^2(\tau)$ to find the diffusion coefficient
$D_r$ and its angular uncertainty $\epsilon_r$.

We find $D_r = 0.311 \pm 0.034$ rad$^2$/s
(Figure~\ref{fig:cylinderresult}), which is larger than that predicted
by the cylinder model ($D'_r = 0.205 \pm 0.008$ rad$^2$/s) but in
quantitative agreement with the spherocylinder model ($D'_r = 0.311
\pm 0.017$ rad$^2$/s). The error obtained from the fits is $\epsilon_r
= 0.0265$ rad, yielding an angular tracking precision of
1.5$^\circ$. Like the translation results, these results suggest that
the rods are modeled well, both hydrodynamically and optically, as
spherocylinders.  Furthermore the tracking precision is better than
that reported for nanorods using reconstructions (approximately
3$^\circ$) and dimers using multisphere superposition solutions
(3.4$^\circ$)~\cite{cheong_rotational_2010, fung_measuring_2011}.

\subsection{Janus particles}
\label{sec:janusresult}

We find that it is more difficult to fit our scattering model to the
Janus particle holograms, as they are only weakly asymmetric.  In
particular, we find that the best-fit polar angle $\theta$ is often
spurious.  There are two local minima in the objective function,
corresponding to two polar angles reflected about the x-y plane, and
two different $z$-coordinates: if we calculate a hologram of a Janus
particle with $\theta = 0$ and $z$ = 8.0 $\mu$m, and we fit to it
using an initial guess of $\theta \ge \pi/2$, the fit converges to
$\theta = \pi$ and $z$ = 7.811 $\mu$m.  We find an $R^2$ = 0.9997 and
a per-pixel $\chi^2$ of 3.8 $\times$ 10$^{-6}$, showing that the
best-fit and original holograms are essentially identical.  These
calculations suggest that errors in the best-fit polar angle are
correlated with those of the $z$-coordinate.  Noise in the hologram
could therefore cause the fitting algorithm to converge to either
local minimum.  Indeed, we estimate our noise floor for the 8-bit
images to be at least an order of magnitude larger at $(1/255)^2 = 1.5
\times 10^{-5}$~\cite{fung_holographic_2013}. Increasing the asymmetry
of the hologram, for example by increasing the aspect ratio or using a
metal-coated Janus particle as in
reference~\cite{anthony_single-particle_2006}, should eliminate this
problem. But for the results shown below, we determine dynamical data
only from the best-fit $x$- and $y$-coordinates and the azimuthal
angle $\psi$.

\subsubsection{Theoretical predictions}

We use the Stokes-Einstein and Stokes-Einstein-Debye
relations~\cite{einstein_investigations_1956, debye_polar_1929} to
model the translational and rotational diffusion of the Janus
particles, which are approximately spherical:
\begin{equation}
\label{eq:stokes_t}
D'_t =k_BT/6\pi\eta a
\end{equation}
\begin{equation}
\label{eq:stokes_r}
D'_r = k_BT/8\pi\eta a^3
\end{equation}

Because the hydrodynamic radii of polymer particles---like the PS core
in our Janus particle---tend to be larger than the radii measured
optically or with electron microscopy due to charged or hairy
surfaces~\cite{gittings_determination_1998, seebergh_evidence_1995},
we do not directly compare the measured diffusion coefficients to
values predicted from the theory.  Instead, we calculate the effective
radius $a_\mathrm{eff}$ from both the measured translation and
rotational diffusion coefficients using Equations~\eqref{eq:stokes_t}
and~\eqref{eq:stokes_r}.  We then compare these values to the best-fit
diameter from the holograms and to each other.

\subsubsection{Translational diffusion}

To determine the translational diffusion coefficients for the Janus
particle, we ignore the translational motion in the direction parallel
to the imaging axis ($z$-axis) because the polar angle $\theta$ has an
uncertainty that affects the best-fit z-position, as discussed
previously.  We treat the Janus particle as a sphere and calculate the
translational diffusion coefficient from the mean-square displacement
projected onto the $x-y$ plane:
\begin{equation}
  \left\langle \Delta {x^2} (\tau)+ \Delta {y^2} (\tau) \right\rangle = 4D_{\perp,\mathrm{Janus}}\tau + 4\epsilon^2
\end{equation}
where $\epsilon$ is the tracking precision.

\begin{figure}[htbp]
\centering
\includegraphics{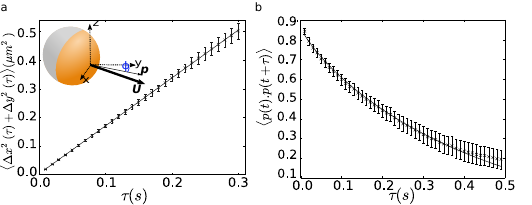}
\caption{\label{fig:janusresult} (a) Measured mean-square
  displacement of the Janus particle in the axes perpendicular to the
  imaging axis. The solid line is a fit to $4D \tau +
 4\epsilon^2$. (b) Measured autocorrelation of the particle's
  projected orientational unit vector $\mathbf{p}$. The solid
  line is a fit to an multi-exponential decay.}
\end{figure}

By fitting the mean-square displacement to a linear model, we obtain
$D_{\perp,\mathrm{Janus}} = 0.419 \pm 0.003$ $\mu$m$^2$/s (see
Figure~\ref{fig:janusresult}). We then use
Equation~\ref{eq:stokes_t} at $21\pm 2$$^\circ$C to obtain an
effective particle radius $a_\mathrm{eff}$ = $524 \pm 15$ nm, which is
larger than the optical radius from holography ($443 \pm 20$ nm for
bare polystyrene, $490 \pm 20$ nm with the TiO$_2$ layer). As
discussed above, the larger hydrodynamic radius is expected for
polymer particles~\cite{gittings_determination_1998,
  seebergh_evidence_1995}. These results are summarized in
Table~\ref{tab:janus_results}.

We obtain a tracking precision of $\epsilon$ = 21.8 nm from the fits
for the directions perpendicular to the imaging axis, which is smaller
than the size of one voxel (approximately 35 nm) and again on par with
the tracking precision of complex scatterers such as spheres in
clusters~\cite{fung_measuring_2011, perry_real-space_2012}.

\begin{table*}
  \centering
    \begin{tabular}{ l l l l}
Quantity & Experiment & Expected & Source \\ \hline \hline

$a_\mathrm{Janus}$ (nm) & $443 \pm 20$ & $450 \pm 50$ & SEM \\
$t$ (nm) & $47 \pm 8$ & 50 & deposition parameters\\
$n_\mathrm{TiO_2}$ & $2.74 \pm 0.23$ & 2.3--2.8 & bulk value\\
$n_\mathrm{PS}$ & $1.581 \pm 0.046$ & $\le1.585$ & bulk value\\
Translational $a_\mathrm{eff}$ (nm) & $524 \pm 15$ & -- \\
Rotational $a_\mathrm{eff}$ (nm) & $523 \pm 6$ & --\\
\end{tabular}
  \caption{\label{tab:janus_results} Measured and expected values of parameters for Janus particles.}
\end{table*}

\subsubsection{Rotational diffusion}
\label{janusrot}

Although the best-fit polar angle $\theta$ is unreliable, we can
extract the rotational diffusion coefficients from the data by
considering only the azimuthal angle $\psi$. To do this, we derive an
expression for the autocorrelation of the projection of the
orientational unit vector onto the $x-y$ plane $\vect{p}(t) =
\cos\psi(t) \,\xhat + \sin\psi(t) \,\yhat$:
\begin{equation}\label{projectionresult}
\left\langle \vect{p}(t)\cdot\vect{p}(t+\tau)\right\rangle =
\frac{1}{4}\sum_{\ell=1}^\infty \frac{2\ell+1}{\ell(\ell+1)}
\left(S_\ell^1 \right)^2 \exp\left[-\ell(\ell+1)D_r\tau \right].
\end{equation}
where $S_\ell^1 \equiv \int_{-1}^{1} P_\ell^1(x)\,dx$ and
$P_\ell^1(x)$ are associated Legendre polynomials of order 1
(see Appendix for further details). This expression is a
multiexponential that depends on $D_r$ and can be evaluated
numerically.

To determine the rotational diffusion coefficient we fit
Equation~\ref{projectionresult} to the data and obtain
$D_{r,\mathrm{Janus}} = 1.15 \pm 0.05$ s$^{-1}$ (see
Figure~\ref{fig:janusresult}). From Equation~\ref{eq:stokes_r} and a
temperature of $21\pm 2$$^\circ$C, we calculate an effective particle
radius of $a_\mathrm{eff} = 523 \pm 6$ nm, in excellent agreement with
the $a_\mathrm{eff} = 524 \pm 15$ nm obtained from translational
motion. These results are summarized in Table~\ref{tab:janus_results}.
The quantitative agreement between the hydrodynamic radii calculated
from the translational and rotational diffusion coefficients
demonstrates the internal consistency of our fitting method, showing
that the technique can effectively track the rotation and translation
of Janus particles in water, despite the small asymmetry of the
particles and holograms.

Although we do not use the $z$-coordinate or the polar angle $\theta$
to obtain these results, they would nonetheless be difficult to obtain
with a traditional optical microscope. The particles diffuse rapidly,
moving more than 5 $\mu$m in $z$ throughout the trajectory, which
might take them too far out of focus during a 2D microscopic
measurement.  Furthermore, the fit to the scattering model allows a
quantitative determination of the azimuthal angle with no calibration
required.  Tracking this angle could be useful in other studies, for
example to determine how quickly a particle orients in response to an
external field or another object.

\section{Conclusions and future work}
We have shown a new technique to measure the 3D translational and
rotational dynamics of colloidal particles.  Our results show that
holographic microscopy can capture the 3D position and orientation of
non-spherical colloidal particles, and that these variables can be
tracked precisely by fitting scattering models based on the discrete
dipole approximation to the measured holograms.  The technique has
high precision and temporal resolution, as evidenced by the
measured rotational diffusion coefficient for the Janus particle,
which is an order of magnitude larger than previously measured 3D
diffusion coefficients of a freely-diffusing particle.

We can address many of the limitations of our approach by improving
our DDA implementation and fitting procedure. We intend to improve the
accuracy of the measurements by antialiasing the voxelation and
incorporating near-field corrections to the scattering calculations,
as described in reference~\cite{dagostino_enhanced_2009}. Spurious
fits might be suppressed with an additional fitting pass that enforces
a physically plausible trajectory, perhaps assisted by a Kalman filter
\cite{fricks_time-domain_2009}.

Because the DDA is applicable to scatterers with arbitrary size,
shape, and refractive index profiles, our technique could be used to
measure the dynamics of a wide variety of particles.  This makes it
suitable for a number of different applications, including
microrheology, measurements of interactions between non-spherical
particles, and fundamental studies of colloidal self-assembly and
bacterial motion.

\section*{Acknowledgments}
We thank Jennifer A. Lewis for helpful discussions, and feedback on
this manuscript.

This work was funded by the Harvard MRSEC through NSF grant
no.~DMR-0820484.  Scattering calculations and hologram fitting were
performed on the Odyssey cluster, managed by the Harvard FAS Sciences
Division Research Computing Group.

SR and IK acknowledge support from the National Science Foundation
through awards CBET-1067501 and CBET-1264550. AW thanks The
University of Sydney Travelling Scholarship for support. TD
acknowledges support from a National Science Foundation Graduate
Research Fellowship.  KC was supported by the AFOSR MURI
award FA9550-12-1-0471.

\appendix
\section{Projection Correlation Function $\langle \vect{p}(t)
\cdot \vect{p}(t + \tau) \rangle$}
In this Appendix, we derive Eq. \ref{projectionresult}, with which we
measure the rotational diffusion coefficient $D_r$ of a Janus particle in
Sec.~\ref{sec:janusresult}.

The isotropic rotational diffusion of a particle can be quantified by
studying the trajectory on the unit sphere of a unit vector $\vect{u}$
fixed to the particle. Computing the autocorrelation $\langle
\vect{u}(t) \cdot \vect{u}(t+\tau)\rangle$, where $\tau$ is a lag
time, from experimental data allows the measurement of the rotational
diffusion coefficient $D_r$. This requires tracking the entire 3D
orientation of the particle.  Here we consider how $D_r$ can be
measured when only a two-dimensional azimuthal projection of
$\vect{u}$ is observed, as is the case for the Janus particles.

In our experiments, we observe the normalized projection of $\vect{u}$
onto the laboratory $x-y$ plane, which is perpendicular to the optical
axis.  In spherical polar coordinates, where the tip of $\vect{u}(t)$
has coordinates $\theta(t)$ and $\psi(t)$ on the unit sphere, this
projection is given by $\vect{p}(t) = \cos\psi(t) \,\xhat +
\sin\psi(t) \,\yhat$.  From the data, we can then compute the
autocorrelation of $\vect{p}(t)$:
\begin{multline}
\langle \vect{p}(t+ \tau) \cdot \vect{p}(t) \rangle =
\langle \cos(\psi(t+\tau)) \cos(\psi(t)) \\+ \sin(\psi(t+\tau))\sin(\psi(t))
\rangle.
\end{multline}
To simplify the notation, we will use primes to denote angles at time
$t + \tau$; the unprimed angles $\theta$ and $\psi$ are at time $t$. Thus,
\begin{equation}\label{eq:correlation_in_angles}
\langle \vect{p}(t+ \tau) \cdot \vect{p}(t) \rangle =
\langle \cos\psi' \cos\psi + \sin\psi'\sin\psi \rangle.
\end{equation}

We show that $D_r$ can be determined from experimental measurements of
$\langle \vect{p}(t+\tau) \cdot \vect{p}(t) \rangle$ by calculating
this autocorrelation for a particle undergoing isotropic rotational
diffusion characterized by $D_r$. We neglect translation-rotation
coupling and therefore ignore the translational diffusion of the
particle. Let $f(\theta,\psi;t)\,d\Omega$ be the probability of
finding $\vect{u}$ in the solid angle $d\Omega$ near $(\theta,\psi)$
at time $t$. The probability density $f$ is governed by a rotational
Fick's law~\cite{berne_dynamic_1976}:
\begin{equation}\label{eq:rot_fick}
\frac{\partial f}{\partial t} = D_r \left(\frac{1}{\sin\theta}
\frac{\partial}{\partial \theta}\left( \sin\theta
\frac{\partial f}{\partial\theta}\right) + \frac{1}{\sin^2\theta}
\frac{\partial^2f}{\partial\psi^2}\right).
\end{equation}
The operator on the right is the Laplacian on the unit sphere. Computing
$\langle \vect{p}(t+\tau) \cdot \vect{p}(t) \rangle$ requires knowing the
transition probability density $K(\theta,\psi,\theta',\psi';\tau)$ for
$\vect{u}$ to move from $(\theta,\psi)$ to $(\theta',\psi')$ after
a lag time $\tau$. If we assume that the distribution of initial
orientations $(\theta,\psi)$ is uniform, such that $f = 1/(4\pi)$,
then using Eq.~\ref{eq:correlation_in_angles} the autocorrelation of
$\vect{p}$ will be given by~\cite{berne_dynamic_1976}
\begin{multline}\label{eq:key_integral}
\langle \vect{p}(t)\cdot \vect{p}(t+\tau)\rangle =\\
\iint (\cos\psi\cos\psi' + \sin\psi\sin\psi')
\frac{K(\theta,\psi,\theta',\psi';\tau)}{4\pi}\,d\Omega\,d\Omega'.
\end{multline}

The transition probability $K$ is given by the probability density
$f(\theta',\psi';\tau)$, governed by Eq.~\ref{eq:rot_fick}, with the
following initial condition:
\begin{equation}\label{eq:intl_condition}
f(\theta',\psi';0) = \frac{\delta(\theta' - \theta) \delta(\psi'-\psi)}
{\sin\theta'}
\end{equation}
where $\delta$ denotes the Dirac delta function. Separation of variables
leads to the following solution for $K$:
\begin{align} \label{eq:k_expansion}
K(\theta,\psi,\theta',\psi';\tau) =
\sum_{\ell=0}^\infty C_{\ell0}P_\ell(\cos\theta')
\exp\left[-\ell(\ell+1)D_r\tau \right] \nonumber \\
 + \sum_{\ell=1}^\infty \sum_{m=1}^\ell \sum_{p=1}^2
C_{\ell m}^{(p)} Y_{\ell m}^{(p)}(\theta',\psi')
\exp\left[-\ell(\ell+1)D_r\tau \right].
\end{align}
Here, $P_\ell(\cos\theta')$ is a Legendre polynomial, and the
$Y_{\ell m}^{(p)}$ are real spherical harmonics~\cite{morse_methods_1953}:
\begin{equation}
Y_{\ell m}^{(p)}(\theta',\psi') \equiv
\begin{cases}
P_{\ell}^m(\cos\theta') \cos m\psi' & \text{if } p = 1 \\
P_{\ell}^m(\cos\theta') \sin m\psi' & \text{if } p = 2.
\end{cases}
\end{equation}
The initial condition in Eq.~\ref{eq:intl_condition}
results in the expansion coefficients being
\begin{equation}
C_{\ell 0} = \frac{2\ell+1}{4\pi}P_\ell (\cos\theta)
\end{equation}
for the azimuthally symmetric ($m=0$) terms and
\begin{equation}
C_{\ell m}^{(p)} = \frac{2\ell+1}{2\pi} \frac{(\ell -m)!}{(\ell+m)!}
Y_{\ell m}^{(p)}(\theta,\psi)
\end{equation}
for the remaining terms.

Consider the first term on the right side of
Eq.~\ref{eq:key_integral} which
contains $\cos\psi\cos\psi'$. By orthogonality of $\cos\psi$ and
$\cos\psi'$, only terms in $K$ with $m=1$ and $p=1$ contribute to the
integral. Integration over $\psi$ and $\psi'$ contributes two
factors of $\pi$, and so we obtain
\begin{align} \hspace{2em}&\hspace{-2em}
\iint \cos\psi\cos\psi'
\frac{K(\theta,\psi,\theta',\psi';\tau)}{4\pi}\,d\Omega\,d\Omega' =
\nonumber \\
& \sum_{\ell=1}^\infty \left( \frac{2\ell+1}{\ell(\ell+1)}
\frac{1}{8\pi^2}\exp\left[-\ell(\ell+1)D_r\tau \right] \pi^2 \right.
\nonumber \\
&\times \left. \int P_\ell^1(\cos\theta)\sin\theta\,d\theta
\int P_\ell^1(\cos\theta')\sin\theta' \,d\theta' \right).
\end{align}
Integrating the remaining term of Eq.~\ref{eq:key_integral} over
$\psi$ and $\psi'$ gives exactly the same result. Defining
\begin{equation}\label{eq:Sl1_defn}
S_\ell^1 \equiv \int_{-1}^{1} P_\ell^1(x)\,dx,
\end{equation}
we obtain Eq.~\ref{projectionresult} in the manuscript:
\begin{equation}\label{eq:main_result}
\langle \vect{p}(t)\cdot\vect{p}(t+\tau)\rangle =
\frac{1}{4}\sum_{\ell=1}^\infty \frac{2\ell+1}{\ell(\ell+1)}
\left(S_\ell^1 \right)^2 \exp\left[-\ell(\ell+1)D_r\tau \right].
\end{equation}
Note that $S_\ell^1 = 0$ for even $\ell$ due to the parity of
$P_\ell^1(x)$. To evaluate our result numerically, we use
DiDonato's recursion relation for $S_\ell^1$~\cite{didonato_recurrence_1982}:
\begin{equation}
S_{\ell+2}^1 = \frac{\ell(\ell+2)}{(\ell+1)(\ell+3)}S_\ell^1.
\end{equation}

Unlike the results obtained by prior workers on this problem, our
solution can be easily computed and used to measure $D_r$ from
experimental data. Saragosti \textit{et al.}  obtain a series
expression equivalent to Eq.~\ref{eq:main_result}, but their solution
contains complicated angular integrals that are left
unevaluated~\cite{saragosti_modeling_2012}. They therefore determine
$D_r$ from a $\ell = 1$ approximation of
Eq.~\ref{eq:main_result}. While the $\ell=1$ term indeed dominates
when $D_r\tau$ is large, the autocorrelation can be measured most
precisely near $\tau = 0$, at which the number of independent angular
displacements is largest. In this regime, the single-exponential
approximation of Saragosti \textit{et al.} fails, and our full
solution is necessary.

\begin{figure}
\centering
\includegraphics{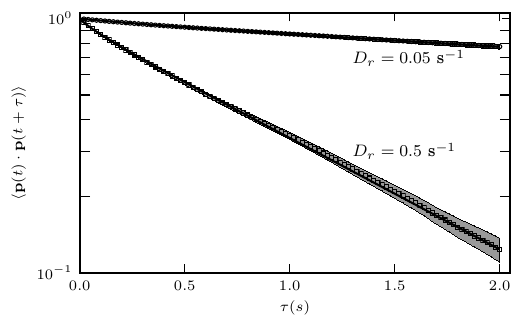}
\caption{\label{fig:projection_simulation} Autocorrelation $\langle
  \vect{p}(t) \cdot \vect{p}(t+\tau)\rangle$ for simulated particle
  undergoing rotational diffusion with $D_r = 0.05$ s$^{-1}$ (open
  circles) and $D_r = 0.5$ s$^{-1}$ (open squares). Solid lines are
  best fits to Eq.~\ref{eq:main_result}. Shaded gray regions denote
  error bars on the autocorrelations, calculated using a block
  decorrelation technique~\cite{flyvbjerg_error_1989}.}
\end{figure}

We verified our result, Eq.~\ref{eq:main_result}, by computing
$\langle\vect{p}(t) \cdot \vect{p}(t+\tau) \rangle$ for simulated
rotational trajectories of particles undergoing rotational diffusion.
We simulated rotational diffusion using the algorithm of Beard and
Schlick~\cite{beard_unbiased_2003}. Figure
\ref{fig:projection_simulation} shows the autocorrelations computed from
simulated trajectories with two different $D_r$ along with fits to
Eq.~\ref{eq:main_result}. We find excellent agreement between the
simulated autocorrelations and best fits to Eq.~\ref{eq:main_result};
the values of $D_r$ determined from the best fits agree with the
simulation input values to 0.5\% or better.

\bibliographystyle{elsarticle-num}
\bibliography{manuscript}







\end{document}